\documentclass[style=bcnpostgrad,entry,preprint]{proceedings}
\renewcommand{\conferencetitle}{Barcelona Postgrad Encounters on Fundamental Physics}
\usepackage[utf8]{inputenc}
\usepackage{subfigure}

\newcommand{\eqend}[1]{\, #1}

\begin{document}

\author{Zacharias Roupas}
\email{roupas@inp.demokritos.gr}
\affiliation{Institute of Nuclear and Particle Physics, N.C.S.R. Demokritos,\\ GR-15310 Athens, Greece}

\title{Gravitational-thermodynamic instabilities of isothermal spheres in dS and AdS}

\abstract{Thermodynamical stability of fluid spheres is studied in the presence of a cosmological constant, both in the Newtonian limit, as well as in General Relativity. In all cases, an increase of the cosmological constant tends to stabilize the system, making asymptotically de Sitter space more thermodynamically stable than anti-de Sitter at the purely classical level. In addition, in the Newtonian case reentrant phase transitions are observed for a positive cosmological constant, due to its repelling property in this case. In General Relativity is studied the case of radiation, for which is found that the critical radius, at which an instability sets in, is always bigger than the black hole radius of the system and furthermore, at some value of the cosmological constant this critical radius hits at the cosmological horizon.} 

\acknowledgements{Many of the results used in this talk were elaborated jointly with Minos Axenides and George Georgiou and appear analytically elsewhere.}

\maketitle

%
%
%
%
%
%

\section{Introduction}

Thermodynamics and stability analysis of self-gravitating gas in Newtonian Gravity is an old subject that dates back at least to the works of Antonov \cite{antonov} and Lynden-Bell \& Wood \cite{bell-wood}. A classic review on the subject is written by Padmanabhan \cite{padman}, while others include Refs. \cite{katz,deVega:2001zk,deVega:2001zj,Chavanis:2001hd}, as well. In General Relativity, thermodynamical stability of fluid spheres is studied by the author in Ref. \cite{roupas}, in which, this talk is, partially, based. The thermodynamics of self-gravitating gas is different from the ordinary thermodynamics, where the interactions are assumed to take place only between neighbouring bodies. Striking differences are the facts that entropy and energy are not extensive quantities, stable configurations with negative specific heat do exist and the thermodynamic ensembles are \textit{not equivalent}. This non-equivalence of ensembles is not regarding the equilibria, but only the stability of these equilibria. 
That is, in gravitational thermodynamics the various thermodynamic ensembles have different stability properties. 

We are interested in the effect of a cosmological constant to the stability of self-gravitating gas both in the Newtonian limit as well as in General Relativity. We shall not be restricted in the assumed current value of the cosmological constant, but we shall consider any arbitrary value. One can think physically about that, as regarding different possible Universes or as regarding the current Universe in possible different epochs in the past for cosmological models with a time varying cosmological term (e.g. decaying vacuum approaches \cite{waga,woodard,Polyakov:2012uc}). Many results presented in this talk appear in Refs. \cite{Axenides:2012bf,agrNPB,agrJPCS}, where is studied the thermodynamical stability in Newtonian Gravity with a cosmological constant in the microcanonical \cite{Axenides:2012bf} and canonical \cite{agrNPB,agrJPCS} ensembles. 

Let us review briefly the results in Newtonian Gravity in the case with no cosmological constant, which for convenience we shall simply call `flat case'. In the original formulation \cite{antonov,bell-wood}, the self-gravitating gas, consisting of dust particles or stars, is assumed to be bound by a spherical, non-insulating and perfectly reflecting wall, so that it is in the microcanonical ensemble. For this system, global entropy maxima do not exist as Antonov have proved. There exist only local entropy maxima that correspond to metastable equilibria. These exist only for $ER > -0.335 GM^2$ where $R$ is the radius of the sphere and $E$, $M$ the energy and mass of the gas, respectively. Thus, there is a minimum energy down to which equilibria do exist and strangely a \textit{maximum} radius up to which equilibria do exist for a system with negative energy (like the vast majority of gravitating systems in nature). That is, for bigger radii than this maximum radius the system has no equilibria and collapses! 
This is usually called `gravothermal catastrophe'. In addition there is another `weaker' instability, associated with equilibria, that are unstable. Whether the system lies on an unstable or stable equilibrium (in the case $ER > -0.335 GM^2$) depends on the  density contrast $\log(\rho_0/\rho_R)$, where $\rho_0$, $\rho_R$ are the densities at the center and the edge, respectively, or equivalently on the temperature. In the canonical ensemble, the system is assumed to lie in a heat bath and the walls are considered to be insulating. For this, it is known \cite{bell-wood,Chavanis:2001hd} that equilibria do exist only for $TR > 0.40 GM$ where $T$ is the temperature of the system. Thus, there is a minimum temperature and a minimum radius down to which equilibria do exist. Note, that this behaviour is more similar to the general relativistic case, where a self-gravitating system collapses for small and not big radii. There is also a weak instability associated with unstable equilibria. Just like before, whether 
the system lies on an unstable or stable equilibrium (in the case $TR > 0.40 GM$) depends on the  density contrast $\log(\rho_0/\rho_R)$ or equivalently on the energy in this case.

We find \cite{Axenides:2012bf} that in the microcanonical ensemble the presence of a positive cosmological constant tends to stabilize the system while a negative cosmological constant tends to destabilize it. We call a positive cosmological constant `dS case', while a negative one `AdS case'. An increase in the cosmological constant $\Lambda$ causes an increase to the critical radius (up to which equilibria do exist) call it $R_A$, a decrease in the critical energy (down to which equilibria do exist) and a decrease in the critical density contrast that triggers the weak instability. In dS case, a reentrant behaviour is observed, since there appears a second critical radius we call $R_{IA}$, bigger than $R_A$, associated with $\Lambda$. At this radius the equilibria are restored. In the canonical ensemble \cite{agrNPB,agrJPCS}, for an increasing $\Lambda$ the critical radius (down to which equilibria do exist) is decreasing, the critical temperature (down to which equilibria do exist) is decreasing and the 
critical density contrast is decreasing as well. In dS case, a reentrant phase transition occurs, since there appears a second lower critical temperature where equilibria are restored.

In General Relativity, the walls are not needed in order for the system to come into an equilibrium. A maximum entropy principle has been studied in Refs. \cite{cocke,swz,Gao:2011hh}. In the case of relativistic radiation it has been proven by Sorkin, Wald and Zhang (SWZ) \cite{swz}  for the specific case of radiation only, that the relativistic equation of hydrostatic equilibrium, namely the Tolman-Oppenheimer-Volkov equation (TOV) can be derived by the extremization of entropy and that the microcanonical thermodynamical stability coincides with the linear dynamical stability. Clarifying and generalizing an argument of Gao \cite{Gao:2011hh}, we show \cite{roupas} that TOV can be deduced thermodynamically by either the microcanonical or the canonical ensemble for any equation of state. Most importantly, we show the equivalence of microcanonical thermodynamical stability with linear dynamical stability for a general equation of state in General Relativity.

To study the effect of the cosmological constant on the microcanonical thermodynamical stability of fluid spheres in General Relativity we consider the fluid to be bounded by spherical non-insulating walls. For simplicity we consider only the case of radiation. We find that the critical radius down to which equilibria do exist is decreasing with increasing $\Lambda$. This radius is bigger than the black hole radius of the system for any value of $\Lambda$. At some $\Lambda$ value the critical radius hits the cosmological horizon. Any sphere with bigger $\Lambda$ is stable since matter outside the cosmological horizon cannot interact with matter inside. Finally, dS tends to stabilize the system while AdS to destabilize it, since the critical ratio $M/R$ is increasing with increasing~$\Lambda$.

\section{Newtonian Gravity}

In the presence of the cosmological constant the Poisson equation becomes \cite{agrNPB}:
\begin{equation}\label{eq:poissonL}
	\nabla^2\phi = 4\pi G\rho - 8\pi G \rho_\Lambda\eqend{,}
\end{equation}
where $\rho_\Lambda$ is the mass density associated with the cosmological constant $\Lambda$ given by
$\rho_\Lambda = \Lambda c^2/8\pi G$.
The cosmological constant acts as a radial harmonic force: $\vec{F} = (8\pi G/3)  \rho_\Lambda r \hat{e}_r$, that is repulsive in dS case ($\rho_\Lambda > 0$) and attractive in AdS case ($\rho_\Lambda < 0$).

We consider a self-gravitating gas of $N$ particles with unity mass inside a spherical shell.  We restrict only to spherical symmetric configurations and work in the mean field approximation, where it is used the Boltzmann entropy
\begin{equation}
	S = -k\int f\ln f \total^3\vec{r} \total^3\vec{p}\eqend{,}
\end{equation} 
that is defined in terms of the one body distribution function $f(\vec{r},\vec{\upsilon})$. In the canonical ensemble is used the Helmholtz free energy $F = E - TS$ or equivalently we work with the Massieu function  $J = -F/T$ that gives 
$J = S-\frac{1}{T}E$. The maximization of $S$ with constant $E$ and the maximization of $J$ with constant $T$ (and constant $M$ in both cases), with respect to perturbations $\delta\rho$, are the same \cite{agrNPB} to first order in $\delta\rho$ , i.e. give the same equilibria, described by the Maxwell-Boltzmann distribution function
\begin{equation}\label{eq:rho}
	f = \left( \frac{\beta}{2\pi}\right)^{\frac{3}{2}}\rho(r)\mathe^{-\frac{1}{2}\beta\upsilon^2}\eqend{,}
\qquad 
	\rho(r) = \rho_0 \mathe^{-\beta(\phi - \phi(0))}\eqend{,}
\end{equation}
where $\phi$ satisfies equation (\ref{eq:poissonL}). Although the two ensembles have the same equilibria, the second variations of $S$ and $J$ are different. A positive $\delta^2S$ or $\delta^2J$ for an equilibrium would signify that this equilibrium is unstable in the microcanonical or the canonical ensemble, respectively, since only local maxima of $S$ or $J$ correspond to stable equilibria.

\begin{figure}[!ht]
\begin{center}
	\subfigure[Newtonian in the microcanonical ensemble]{ \label{fig:Rcr_N}\includegraphics[height=6cm]{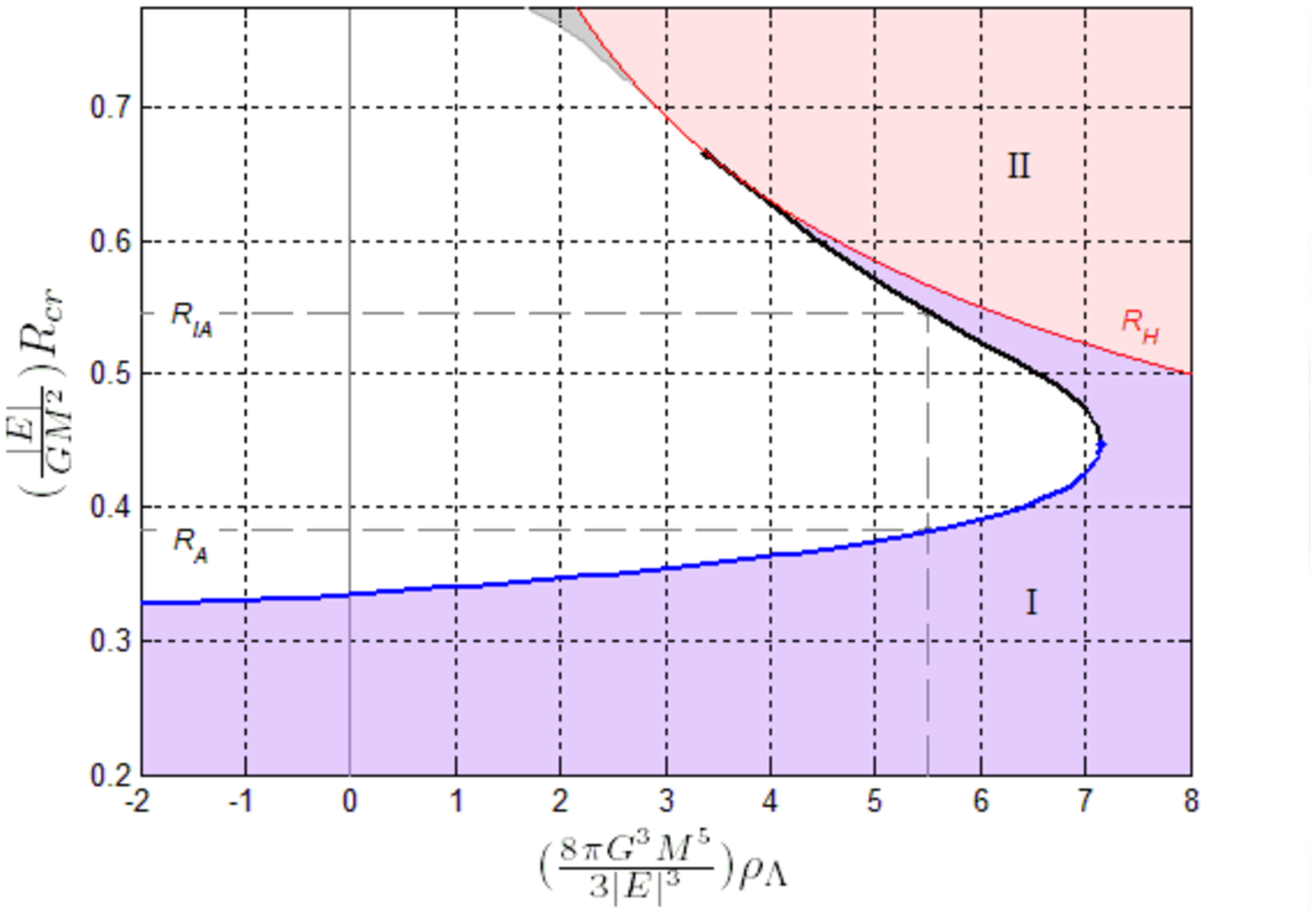} }
	\subfigure[Schwarzschild-dS]{ \label{fig:hor_SdS}\includegraphics[height=5.7cm]{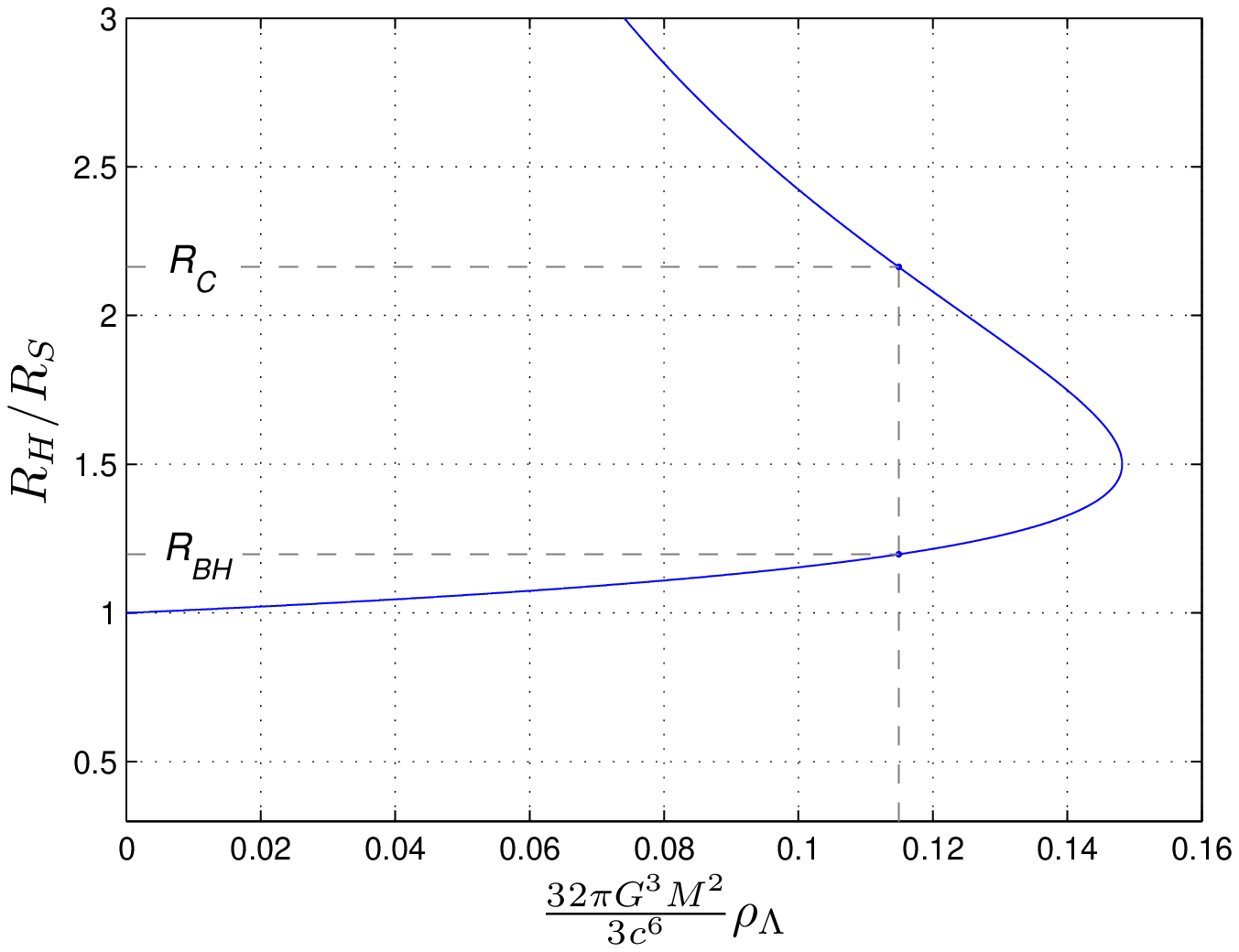} }
\caption{ \small On the left, the critical radius, in the Newtonian self-gravitating gas, versus $\rho_\Lambda$ for fixed $E$, $M$ in the microcanonical ensemble. There exist no equilibria in the unshaded region. With $R_H$ is denoted the radius of the homogeneous solution. On the right, the horizons of the Schwarzschild-dS space versus the cosmological constant for a fixed mass. The horizon radius $R_H$ is measured in units of the Schwarzschild radius $R_S = \frac{2GM}{c^2}$. For a given cosmological constant there are two horizons: the black hole horizon $R_{BH}$ and the cosmological horizon $R_C$. The similarity with Figure \ref{fig:Rcr_N} is striking! Both figures appeared originally in Ref. \cite{Axenides:2012bf}.
\label{fig:Rcr}}
\end{center}
\end{figure}

Introducing the dimensionless variables $y = \beta(\phi - \phi (0))$, $x = r\sqrt{4\pi G \rho_0\beta}$ and $\lambda = 2\rho_\Lambda/\rho_0$, and using equation (\ref{eq:rho}), equation (\ref{eq:poissonL}) becomes:
\begin{equation}\label{eq:emdenL}
	\frac{1}{x^2}\frac{\total}{\total x}\left( x^2\frac{\total}{\total x}y\right) = \mathe^{-y} - \lambda\eqend{,}
\end{equation}
called the Emden-$\Lambda$ equation. Let us call $z = R\sqrt{4\pi G \rho_0\beta}$ the value of $x$ at $R$. In order to generate the series of equilibria needed to study the stability of the system, the Emden-$\Lambda$ equation has to be solved with initial conditions $y(0)=y'(0)=0$, keeping $M$ constant and for various values of the parameters $\rho_\Lambda$, $\beta$, $\rho_0$. This is a rather complicated problem, since, unlike $\Lambda = 0$ case, while solving for various $z$, mass is not automatically preserved, because of the mass scale $M_\Lambda = \rho_\Lambda\frac{4}{3}\pi R^3$ that $\Lambda$ introduces. A suitable $\lambda$ value has to be chosen at each $z$. We define the dimensionless mass 
\begin{equation}\label{eq:mDEF}
	m \equiv \frac{M}{2M_\Lambda} = \frac{3}{8\pi}\frac{M}{\rho_\Lambda R^3} = \frac{\bar{\rho}}{2\rho_\Lambda}\eqend{,}
\end{equation}
where $\bar{\rho}$ is the mean density of matter. Calling $z = R\sqrt{4\pi G \rho_0\beta}$ the value of $x$ at $R$, $m$ can also be written as $m = 3 B/\lambda z^2$, where 
\begin{equation}
	B = \frac{GM\beta}{R}
\end{equation}
is the dimensionless inverse temperature. 
It can be calculated by integrating the Emden-$\Lambda$ equation, to get:
\begin{equation}\label{eq:beta}
	B(z) = z y'(z) + \frac{1}{3}\lambda z^2\eqend{.}
\end{equation}
We developed an algorithm to solve equation (\ref{eq:emdenL}) for various values of $\lambda$, $z$ keeping $m$ fixed.
From equation (\ref{eq:mDEF}) it is clear that solving for various fixed $m$ can be interpreted as solving for various $\rho_\Lambda$ and/or $R$ for a fixed $M$.

The dimensionless energy is defined as $Q = \frac{ER}{GM^2}$. It can be calculated by use of the virial theorem $2K + U_N - 2U_\Lambda = 3PV$, where $K$, $U_N$, $U_\Lambda$, $P$ and $V$ are the kinetic energy, the Newtonian potential energy, the cosmological potential energy, the pressure and the volume, respectively. The dimensionless energy $Q$ is found \cite{Axenides:2012bf,agrNPB} to be:
\begin{equation}\label{eq:Eenergy}
	Q(z) = \frac{z^2 \mathe^{-y}}{B^2} - \frac{3}{2B} - \frac{\lambda}{2B^2 z}\int{ \mathe^{-y}   x^4 \total x}\eqend{.}
\end{equation}

\begin{figure}[!ht]
\begin{center}
	\subfigure[Critical radius in the canonical ensemble]{ \label{fig:Rcr_C}\includegraphics[height=6cm]{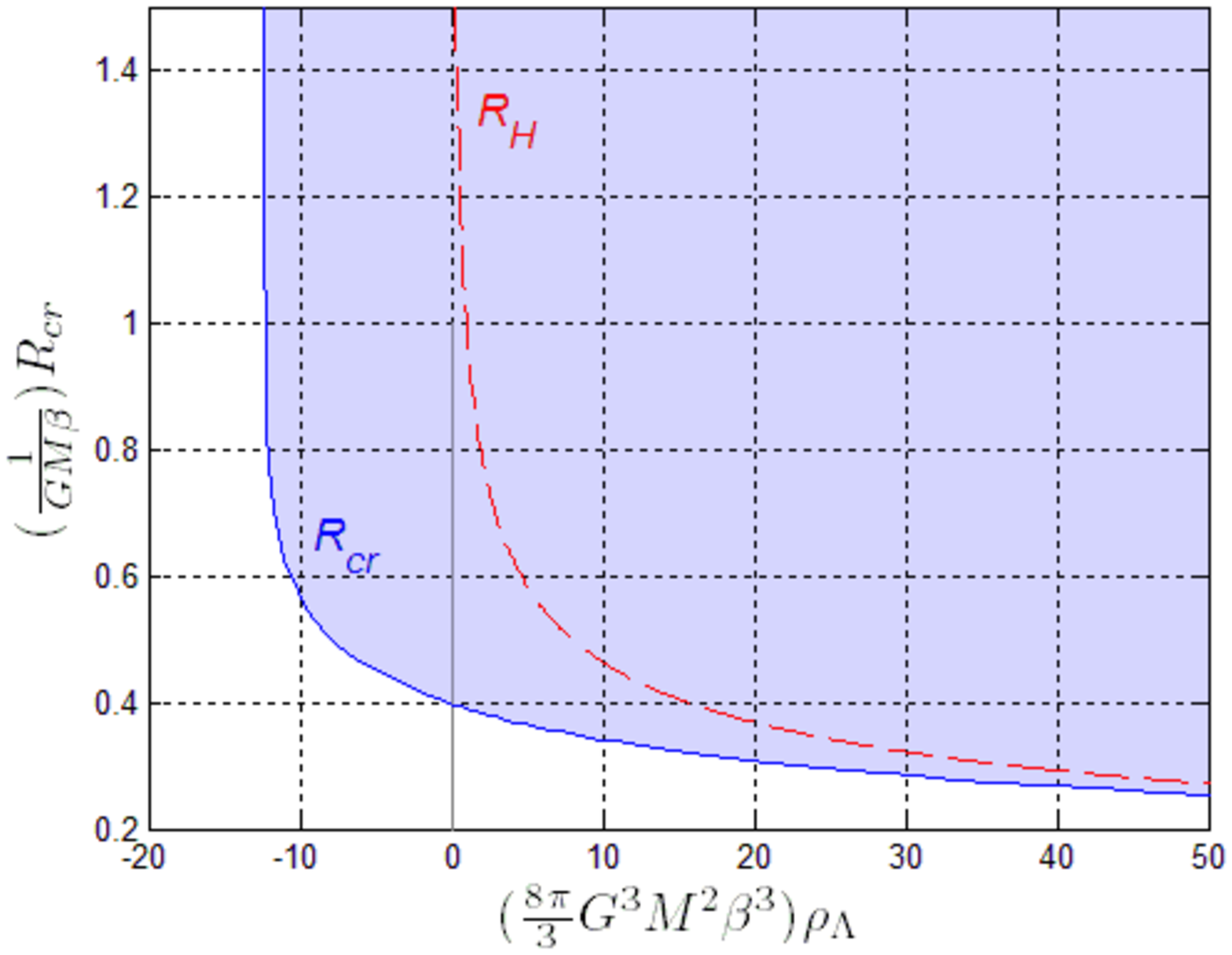} }
	\subfigure[Critical temperature in the canonical ensemble]{ \label{fig:Tcr_C}\includegraphics[height=6cm]{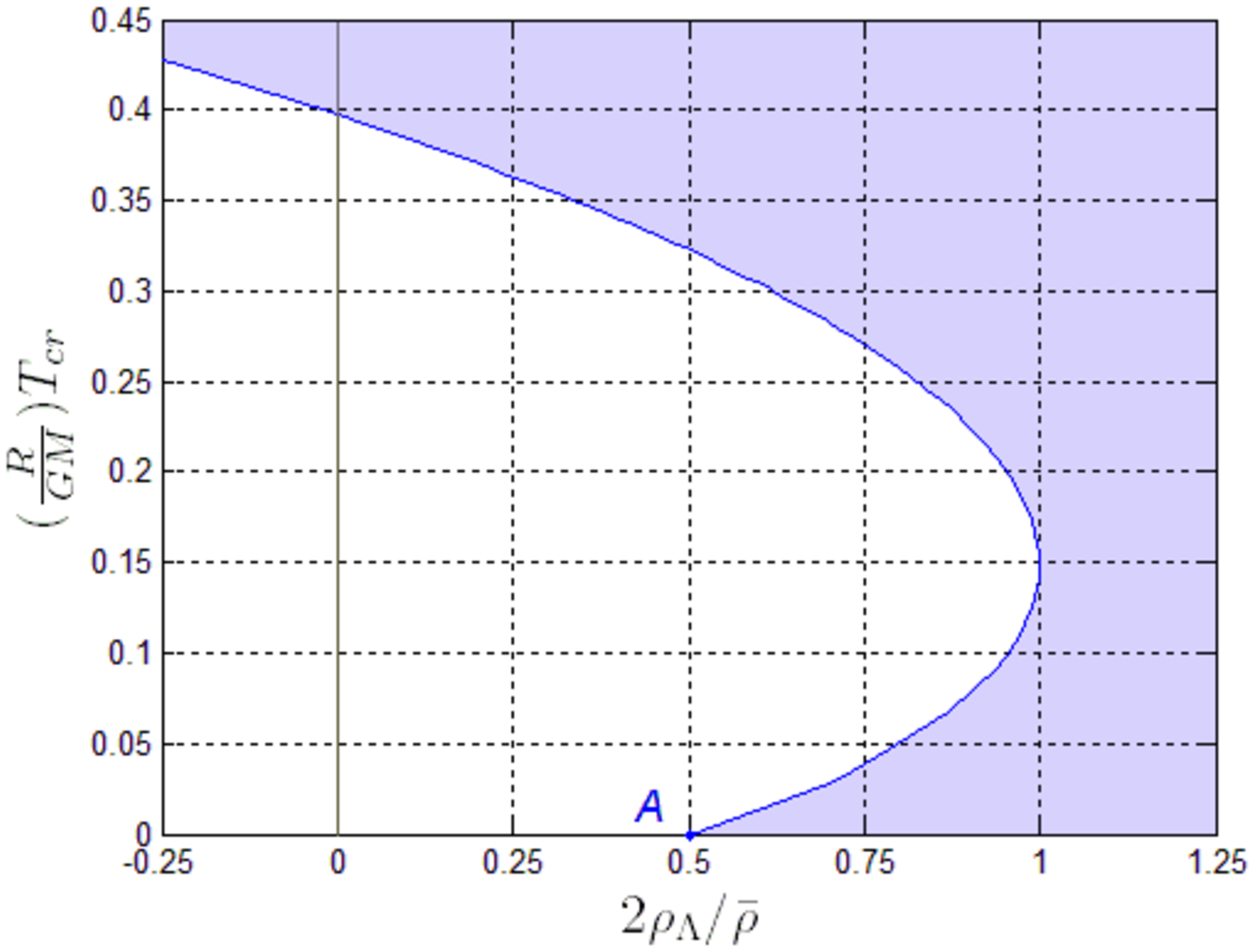} }
\caption{\small On the left, the critical radius versus $\rho_\Lambda$ for fixed $\beta$, $M$ in the canonical ensemble. There exist no equilibria in the unshaded region. With $R_H$ is denoted the radius of the homogeneous solution. On the right, the critical temperature versus $\rho_\Lambda$ for fixed $M$, $R$ in the canonical ensemble, where $\bar{\rho}$ is the mean density of matter. In the unshaded region there exist no equilibria. This behaviour indicates a reentrant phase transition. Both figures appeared originally in Ref. \cite{agrNPB}.
\label{fig:Canon}}
\end{center}
\end{figure}

According to Poincar\'e's theorem \cite{poincare,katz1}, the maximum of $Q(z)$ in a series of equilibria $z$, is the turning point of stability in the microcanonical ensemble, while the maximum of $B(z)$ is the turning point of stability in the canonical ensemble. At this equilibrium points a stable branch of equilibria becomes unstable. Whether the system lies on a stable or unstable equilibrium depends then on the density contrast $\log(\rho_0/\rho_R)$. The very same points mark the marginal conditions for which equilibrium states to exist, since for any $Q$ or $B$, respectively, above these points there are no equilibria at all.

In dS case there are found \textit{multiple} series of equilibria. For some of them the density is increasing towards the edge and others have periodic condensations, like a core-halo structure with one or even more haloes. In addition there appears a homogeneous solution at the radius $R_H = (3M/8\pi \rho_\Lambda)^{1/3}$, that is the equivalent to Einstein's static Universe in the Newtonian limit. This solution is stable for temperatures higher than the critical value $T_h = GM/6.73R_H$. In Figure \ref{fig:Rcr_N}, in the region I are solutions with $R<R_H$ and monotonically decreasing density towards the edge of the sphere, in region II there are solutions with $R>R_H$ and monotonically increasing density as well as solutions with periodic condensations. In the small gray area there are solutions with $R<R_H$ and periodic condensations. In the same figure, it is evident a reentrant behaviour in the critical radius. At the radius $R_{IA}$, equilibria are restored. A mysterious similarity with the 
relativistic Schwarzschild-dS space, that is an empty space, is obvious in Figure \ref{fig:hor_SdS}. It seems that the reentrant behaviour we observe is the Newtonian analogue of the two horizons in Schwarzschild-dS space. In this analogy, the Newtonian radius $R_N = GM^2/|E|$ is the analogue of the Schwarzschild radius $R_S = 2GM/c^2$. Note that Figure \ref{fig:Rcr_N} is produced by rather complex numerical calculations, while in Figure \ref{fig:hor_SdS} are drawn just the real roots of a simple third order algebraic polynomial.

In the canonical ensemble the system becomes unstable for radii \emph{smaller} than a critical radius. In contrast to the microcanonical ensemble, this is a much more intuitively expected behaviour, similar to the one in General Relativity. The reason of the complete inversion of the region of instability in the two ensembles, lies on the the fact that in the microcanonical ensemble, the pressure gradient will increase during a compression of the system because of the increase in temperature, while in the canonical ensemble, the heat bath will absorb the energy and the pressure gradient will decrease, becoming unable to balance Gravity. In Figure \ref{fig:Rcr_C} is shown the critical radius in the canonical ensemble with respect to the cosmological constant. An increase in the cosmological constant causes a decrease to the critical radius, enlarging the region of stability. In AdS case, we see that beyond some $\Lambda$ value, it is impossible to retain any equilibrium. In Figure \ref{fig:Tcr_C} we observe a 
reentrant phase transition in the critical temperature. As the cosmological constant is increasing the critical temperature, down to which equilibria exist, is decreasing. However, in dS, above the value $\rho_\Lambda^\text{marg.} = \bar{\rho}/4$ a further decrease in the temperature restores the equilibria beyond some second critical temperature. That happens because of the repelling character of positive $\Lambda$, which for low temperature can balance gravity. The value $\rho_\Lambda^\text{marg.}$ is this value of the cosmological constant that can balance gravity point by point at zero temperature, i.e. the point $A$ in Figure \ref{fig:Tcr_C} is a static \textit{mechanical} equilibrium. The value $\rho_\Lambda^\text{marg.}$ is analytically calculated in \cite{Axenides:2012bf}.

Finally, we stress out that an increase of the cosmological constant tends to stabilize the system in both ensembles. This is evident in Figures \ref{fig:Rcr_N}, \ref{fig:Rcr_C}, \ref{fig:Tcr_C} where one can see that the region of stability is increasing as $\rho_\Lambda$ increases.

\section{General Relativity}

\begin{figure}[!ht]
\begin{center}
	\subfigure[Ratio $M/R$ for $q=1/3$]{ \label{fig:MY_RAD}\includegraphics[scale=0.53]{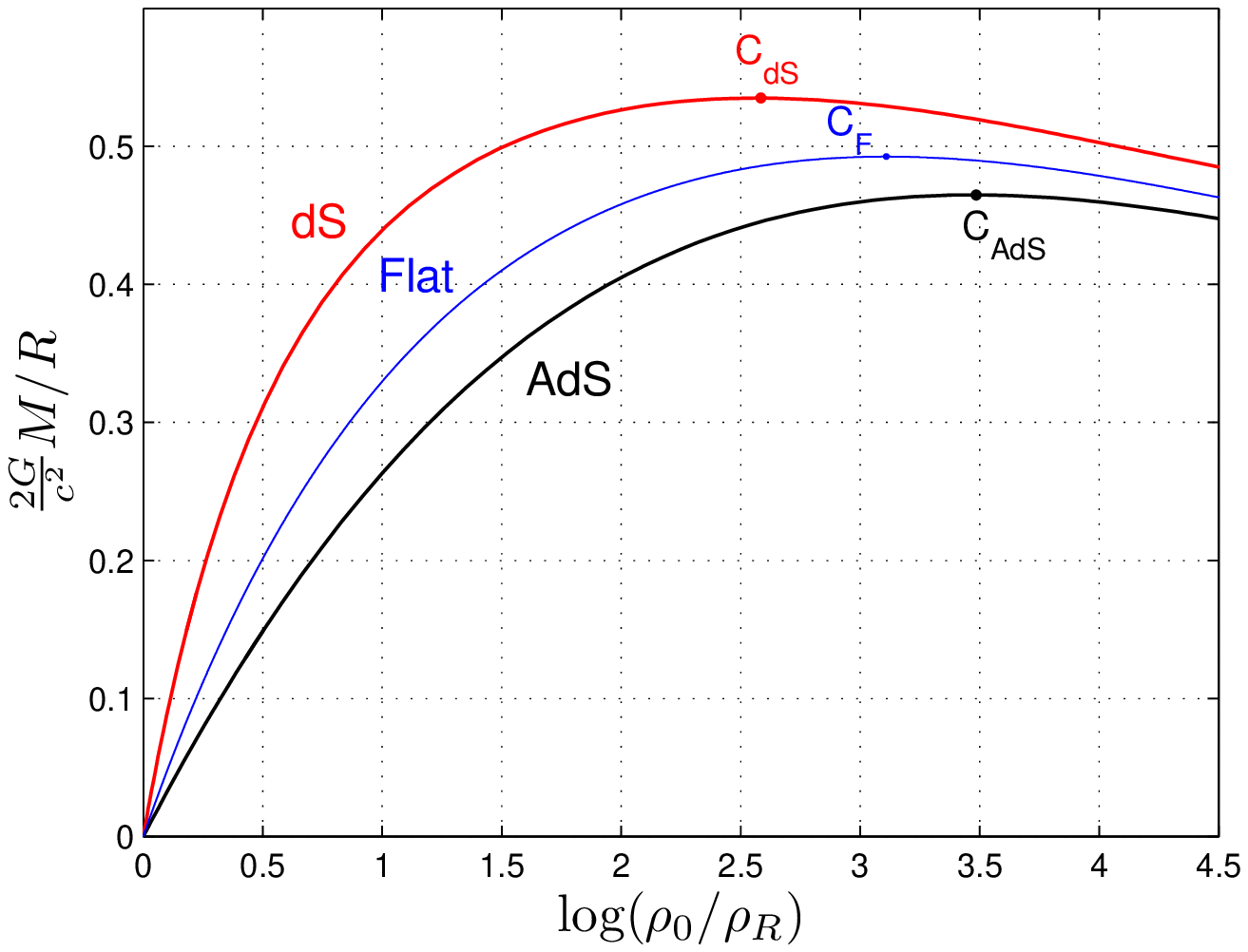} }
	\subfigure[Critical radius for $q=1/3$]{ \label{fig:Rcr_R}\includegraphics[scale=0.53]{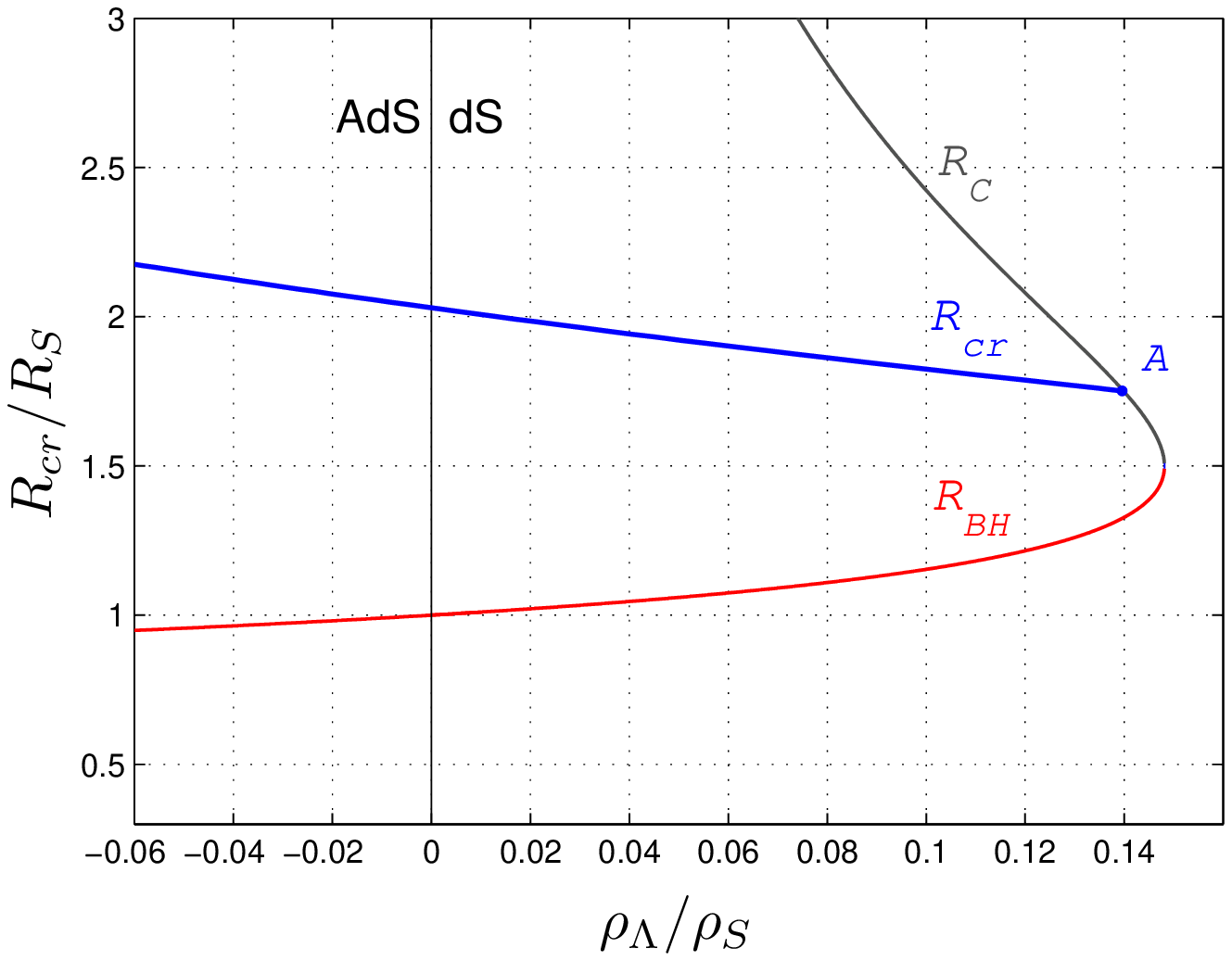} } 
\caption{\small On the left, the series of equilibria expressed as the ratio $M/R$ with respect to the density contrast $\log(\rho_0/\rho_R)$ for radiation ($q=1/3$), for asymptotically flat, de Sitter and anti-de Sitter cases. The points $C_i$ are turning points of stability, i.e. at their left side the equilibria are stable, while at their right side, unstable. At each case above $C_i$ there are no equilibria at all. On the right, the critical radius $R_{cr}$ with respect to the cosmological constant $\rho_\Lambda$ in Schwarzschild units $R_S = 2GM/c^2$, $\rho_S = 3M/4\pi R_S^3$ for a fixed mass $M$. The spheres of radius $R < R_{cr}$ are strongly thermodynamically unstable, i.e. there exist no equilibria at all. At point $A$, the critical radius hits the cosmological horizon. Figure \ref{fig:Rcr_R} appeared originally in Ref. \cite{roupas}.
\label{fig:Rel}}
\end{center}
\end{figure}

In General Relativity, a spherically symmetric, static perfect fluid obeys the Tolman-Oppen\-heimer-Volkov (TOV) equation:
\begin{equation}\label{eq:TOV2}
	p' = -\left( \frac{p}{c^2} + \rho \right)\left( \frac{Gm(r)}{r^2} + 4\pi G \frac{p}{c^2}r - \frac{8\pi G}{3}\rho_\Lambda r\right)
	\left( 1 - \frac{2Gm(r)}{rc^2} - \frac{8\pi G}{3 c^2}\rho_\Lambda r^2\right)^{-1}\eqend{,}
\end{equation}
with $m' = 4\pi r^2\rho(r)$, where $p$ is the pressure of the fluid and $m(r)$ the total mass-energy (including the one of the gravitational field) included inside $r$. This equation expresses the hydrostatic equilibrium in General Relativity and is derived by the Einstein's equations. In Ref. \cite{roupas}, we showed following \cite{swz,Gao:2011hh} that it can also be derived by the extremization of entropy in the microcanonical ensemble or the extremization of free energy in the canonical ensemble. Most importantly, we showed that the second variation of entropy gives the same condition for stability with radial variations to first order in Einstein's equations for a general equation of state. Thus, is evident the equivalence between microcanonical thermodynamical stability and linear dynamical stability. We observed in the previous chapter that the Newtonian canonical ensemble, behaves similar to a relativistic system. We will see in the followings, even more evidence, that the microcanonical ensemble in 
General Relativity becomes the canonical ensemble in the Newtonian limit, hinting at the presence of an implicit heat bath in General Relativity.

For a linear equation of state
\[
	p = q\rho c^2
\]
and introducing the dimensionless variables 
\begin{equation}\label{eq:cvar}
\begin{split}
	\rho &= \rho_0 \mathe^{-y}\; , \;\; x = r \sqrt{4\pi G \rho_0\frac{q+1}{qc^2}}\; ,\;\;
	\lambda = \frac{2\rho_\Lambda}{\rho_0} \; , \\
	\mu  (x) &= \frac{1}{4\pi \rho_0}\left( 4\pi G\rho_0 \frac{q+1}{qc^2} \right)^{\frac{3}{2}}m(r)\; , \;\; M = m(R)
	\; ,\;\; z = x(R) ,
\end{split}
\end{equation}
the TOV equation (\ref{eq:TOV2}) becomes
\begin{equation}\label{eq:TOV_ND}
\begin{split}
	\frac{\total y}{\total x} &= \left( \frac{\mu}{x^2} + q x \mathe^{-y} - \frac{\lambda}{3} x\right) 
		\left( 1 - 2\frac{q}{q+1}\frac{\mu}{x} - \frac{\lambda}{3}\frac{q}{q+1} x^2 \right)^{-1} \eqend{,}\\
	\frac{\total\mu}{\total x} &= x^2 \mathe^{-y}\eqend{.}
\end{split}
\end{equation}
The Emden-$\Lambda$ equation (\ref{eq:emdenL}) is the limit of the above equation for dust matter $q\rightarrow 0$. Just like in the Newtonian case we define the dimensionless mass
$
	\tilde{m} \equiv \frac{M}{2M_\Lambda} = \frac{3}{8\pi}\frac{M}{\rho_\Lambda R^3} = \frac{3\mu}{\lambda z^3}
$
and develop a computer code to solve equation (\ref{eq:TOV_ND}) keeping $\tilde{m}$ fixed. The dimensionless energy becomes now
\begin{equation}\label{eq:Q2}
	Q \equiv \frac{2GM}{Rc^2} = \frac{2\mu}{z}\frac{q}{q+1}   = \frac{2q}{q+1}\frac{zy'(z)-qz^2 \mathe^{-y(z)} - \frac{\lambda}{3}z^2\left(\frac{q}{q+1}zy'(z) + 1\right)}{2\frac{q}{q+1}zy'(z) + 1}\eqend{.}
\end{equation}
Weinberg (see p. 305 in Ref. \cite{weinberg}) has proven that the maxima of $M$ coincide with the maxima of the baryon number $N$ in a series of equilibria, for systems with constant chemical composition and constant entropy per baryon, like systems with a linear equation of state. The maxima of $M$ just like the Newtonian case are turning points of stability for the microcanonical ensemble, in accordance with the Poincar\'e's theorem. So, in our case (linear equation of state) the maxima of $N$ are turning points, as well. We use the dimensionless baryon number
\begin{equation}
\begin{split}
	B(z) &\equiv \frac{N}{N_*} =  \frac{1}{z^{\frac{3q+1}{q+1}}}\int_0^z x^2 \mathe^{-\frac{y}{q+1}} \left(1 - 2\frac{q}{q+1}\frac{\mu}{x}  - \frac{\lambda}{3}\frac{q}{q+1}x^2 \right)^{-\frac{1}{2}} \total x 
	\; , \\
	N_* &= 4\pi R^3 \left(\frac{1}{4\pi G K R^2}\frac{q^2c^4}{q+1} \right)^{\frac{1}{q+1}} ,
\end{split}
\end{equation}
where $K$ is the constant that enters in the polytropic equation $p = K n^{q+1}$ with $n$ the baryon number density. The Newtonian limit $q\to 0$ of $B$ is the quantity 
\[
	B \overset{q\to 0}{\longrightarrow} \frac{GM\beta}{R}\eqend{,}
\]
that is exactly the dimensionless temperature we used in the Newtonian limit, that controls canonical stability. Since as we have seen $B$ controls microcanonical stability in relativity, we deduce that the Newtonian limit of the microcanonical ensemble is the canonical ensemble! This conclusion is supported by even stronger evidence; (a) in \cite{roupas} is proven that the condition for stability in the microcanonical ensemble in General Relativity, becomes the condition for stability in the Newtonian canonical ensemble for non-relativistic dust matter (b) the Newtonian canonical ensemble behaves qualitatively similar to relativistic systems.

Regarding the effect of the cosmological constant on the stability of the system, we focus in the case of radiation, i.e. $q=1/3$. In Figure \ref{fig:Rcr_R} we see how the critical radius, down to which equilibria do exist, is changing with respect to the cosmological constant. The system becomes unstable long before it reaches its black hole radius for any value of the cosmological constant. At some value, it equals the cosmological horizon. Any bigger sphere can be regarded stable, since matter outside the horizon cannot interact with matter inside.

In Figure \ref{fig:MY_RAD} is plotted the ratio $M/R$ of with respect to the density contrast for various equilibria. The maximum point is a turning point of stability and in the same time the marginal point for which equilibria do exist. Since this point goes to bigger values as $\Lambda$ increases we conclude that an increase in the cosmological constant tends to stabilize the system just like in the Newtonian limit.


\end{document}